# A dynamic over games drives selfish agents to win-win outcomes

Seth Frey[1], Curtis Atkisson[2],
[1] Department of Communication, UC Davis
[2] Department of Anthropology, UC Davis

Direct correspondence to Seth Frey, sfrey@ucdavis.edu

**Abstract**

Although individual games are normally used to model isolated interaction types, larger spaces of games can be helpful for modeling the processes by which such interactions change. We introduce a framework for modeling how individuals change the games they are placed in, a process we term "institutional evolution," to contrast with "behavioral evolution," which occurs within- rather than between-game. Starting at random locations in an abstract game space, agents trace trajectories through the space by repeatedly navigating to "neighboring" games, until they converge on "attractor" games that they prefer to all others. Agents choose between games on the basis of "institutional preferences" that define between-game comparisons in terms of game-level features such as fairness and efficiency. We then characterize the attractors of institutional evolutionary over three types of game theoretic agent: the absolute fitness maximizing agent of economic game theory (who is indifferent to inequality), the relative fitness maximizing agent of evolutionary game theory (who maximizes inequality), and the relative group fitness maximizing agent of multi-level/group selection theory (who minimizes inequality). Computing institutional change trajectories over the space of two-player ranked-outcome games, we find that the institutional evolutionary process leads to very different attractors depending on the agent. While "win-win" games account for 25% of games in the space we examine, this proportion is much different among attractor games, either 50%, 0%, or 100% depending on whether agents are economic, evolutionary, or group-level. The result for economic agents is especially interesting: although they are indifferent to the fairness of the games they choose between, their attractors over-represent fairness by 100% relative to baseline, apparently because fairness happens to co-occur with the self-serving features these agents do prefer. We thus present institutional evolution as a mechanism for encouraging the spontaneous emergence of cooperation among inherently selfish agents. We then investigate the sensitivity of these findings to the other two contexts, and to games of more than two players, and we discuss the implications for the emergence of conventions, social systems with large populations, and the mechanisms of major evolutionary transitions. This work provides a flexible, testable formalism for capturing institutional evolutionary process, and for modeling the interdependencies of behavioral and institutional evolutionary processes.

`

**Introduction**

Evolutionary game theory has proven valuable for the analysis of cooperation in a wide variety of biological and social systems, and researchers in the area are increasingly characterizing the conditions necessary for fostering cooperation spontaneously. However, these results continue to treat the games agents play as fixed, disregarding the fact that agents in many game-like settings have incremental influence over the incentive structures they face, and that agents may adjust the games they play toward certain payoff structures. Across the animal world, formalisms allowing agents to change a game's payoffs have provided parsimonious models of the intricacies of sexual selection (1), interactions with resource systems (2), and the emergence of diversity (3). And in the human world specifically, incremental changes to game structures offer a rich model of institutional change at the human level, with findings on preferences for fair punishment (4), on fairness/efficiency tradeoffs (5, 6), and negotiation processes (7). Linking within- and between- game behavior and preferences allows us to study emergent complexity in settings ranging from governance institutions to behavioral ecology.

Here we present a flexible formalism for studying the interactions of within-game "behavioral evolution" — the familiar purview of game theory — with between-game "institutional evolution." Treating normal-form two-by-two games as toy institutions, we represent institutional change dynamics in terms of trajectories over "neighboring" games, games that differ minimally in payoffs. By modeling institutional evolutionary dynamics explicitly, we provide a tractable formalism for generalizing beyond static models of institution formation to testing theories of institutional change. This allows us to move beyond the simplifying assumption in institutional analysis, tacit in theories of human institutional development like Rawl's original position (8), that a planned institution is realized perfectly and remains stable, never deviating in practice from its design.

**Background**

In our framework, the selection pressures driving institutional evolution are players' institutional preferences—the values and qualities agents look for in a social system. Institutional preferences fit within a broader academic interest in human preferences between social constructions such as games, culture, norms, and language (9–11). Institutional preferences have attracted specific interest with theories such as Binmore's, that the processes of cultural evolution select for institutions with the features of stability, efficiency, and fairness, in that order (12). Illustrating the potential for applications to policy, researchers have elicited communities preferences for the features of local natural resource governance institutions (13, 14).

Research on institutional preferences and behavior over game spaces has emerged independently in several disciplines, and efforts to explore larger sets of games are gaining recognition, both in theoretical and experimental work (15, 16). However, theoretical interest in game spaces has put more focus on static comparisons between the games in a space than dynamics across them. Previous work is primarily concerned with behavioral rather than institutional dynamics, in that it leverages large game spaces to catalog the variety of within-game dynamics (17–21), rather than defining preferences over games or trajectories through them.

Among attempts to explore large game spaces, one space in particular, the Topology of Games, has attracted broad formal attempts to taxonomize or otherwise compare behavior across a range of games (22, 23) (see Fig 1a for a sample of the variety of games in the space, and Figs. 2a, S1 for 2D representations of the space ). The space was first organized as a taxonomy by Rapaport (24). Its simplicity and structure make it an ideal substrate for modeling the processes of institutional evolution (25) (Fig 1b).

**Institutional evolution**

We introduce a framework for modeling game change as a trajectory through a space of economic games. In our formalism, agents traverse a lattice of games linked by similarity. They do so in a hill climbing process that optimizes over desirable game features such as stability, efficiency, predictability, etc. To specify this framework, we define a space, distances over it, and the elements of dynamics through it.

<u>Institution space.</u> In the most general terms, we understand institutional evolution as a process in which a set of agents experiences an institution, changes it incrementally in line with their preferences, experiences it again, and so on (Fig. 1b). The first challenge in making this picture concrete is to find a space of social systems that is rich enough to capture a range of human exchange patterns, but simple enough to remain tractable.

We begin with the Topology of Games (22) a space of social systems defined in terms of the two-choice ordinal (ranked-outcome) normal-form games, an arrangement of the 144 unique ways that two agents can assign their own strict rankings over four outcomes (Figs 2a, S1). This space, including extension to *n*-person games in this paper, has several attractive properties. It is simple, composed of the most elementary class of economic game, and amenable to counting. It is also rich: games in the space represent a broad array of social situations (Figs. 1a, 2d). The two-player space includes many of the most famous economic games, such as Prisoner's Dilemma and Chicken, as well as social situations that have traditionally attracted little academic interest, such as no-conflict and win-win games in which individual's choices lead non-strategically to outcomes that benefit all (26). As mundane as these "non-game" games are, their value is clear in the fact that most of our daily social exchanges are similarly mundane. Overall, the space parsimoniously captures an impressive variety of interdependence patterns in human interactions (24, 27).

A major source of appeal for the space is its amenability to formal combinatoric approaches. There are 144 games in this space. 3/4 have exactly 1 pure strategy Nash equilibrium, 1/8 have 0, and 1/8 have two (Fib 2b). As population increases beyond 2, the number of games grows super exponentially, while the count of expected Nash equilibria per game grows more modestly, approaching a standard Poisson distribution, with a game's chances of having 0 or 1 equilibria equal to $e^{-1}$~37% and the remaining quarter have 2 or more (28, 29). Like the number of games, the number of neighbors per game also explodes, following
$$(2^n)^{1-n}$$
where n is the number of players in a game of size
$$\underbrace{2 \times 2 \times \cdots \times 2}_{n}$$

This explosion in the number of neighbors per game implies an even larger explosion in the number of shortest paths between pairs of games, such that simple local strategies like hill climbing can reliably find global optima (30, 31). In a sense, the increasingly convoluted topology of this space, however intriguing, probably does not meaningfully constrain trajectories through it, especially as n increases.

<u>Distance</u>. With the set of games in place, it is possible to introduce a simple conception of distance. We start by restricting our attention to incremental institutional change—trajectories occur over "neighboring" games. Translated to game space, two games are immediate neighbors if they differ only minimally in payoffs (Fig. 2c). Specifically, two games are neighbors if the only difference between them are the locations of a 1 and a 2 ranking, a 2 and a 3, or a 3 and a 4: the Stag Hunt neighbors the Win-Win game because swapping the locations of a 1 payoff and a 2 payoffs will turn the former game into the latter (Fig. 1b shows one possible trajectory from the Prisoner's Dilemma to a Win-Win game).

<u>A dynamic over game space</u>. Given a space and metric over it, we can begin to specify dynamics. Agents alternative playing the current game (on the basis of within-game rules) and choosing which neighboring game to evolve to (on the basis of between-game rules). A dynamic in this framework is thus specified in three parts: the definition of player behavior within a game, the definition of a players' "institutional preferences" for selecting between a game and its neighbors, and the rules for aggregating all agents' game preferences into a single choice. Repeated as stages, a trajectory is produced by repeatedly cycling through the steps of playing a game, eliciting preferences among neighboring games, and selecting a game from the preferences produced. A trajectory has terminated in an attractor game when no neighboring game is preferred to the current one.

<u>The self-interested dynamic</u>. Given this framework, we define a dynamic based on rational, self-interested agents who change the games they play with an eye to institutionalizing their profits and position. This generalizes the absolute fitness maximizing agent of economic game theory. Where such artificially selfish agents converge on prosocial outcomes, more realistic agents are at least as likely to do the same.

Within-game, agents in this self-interested dynamic play rationally, selecting unique pure-strategy Nash equilibria when they exist, and mixed-strategy equilibria otherwise, randomizing over equilibria when several of one type exist.

Across games, a player's institutional preferences define their trajectory. Agents in the self-interested dynamic prefer games that are stable, predictable, and efficient; they prefer a game with a Nash equilibrium that is unique (stable), that is in pure strategies (predictable), and that includes the focal player's top-ranked outcome (efficient). This agent has no social preferences: given two games that are equally stable, predictable, and efficient, players are indifferent as to which most or least benefit others.

The self-interested dynamic's aggregation rule is simple and consistent with a rational agent working to consolidate a beneficial position. The player with greater earnings after the first randomly selected game becomes the focal player choosing subsequent games to move to. Thus, power within a system confers power over it. If the first game results in tied payoffs for two players, the tie is broken randomly. Assuming that the player most recently in control of the dynamic has a small advantage, subsequent ties break in favor of the previous focal player.

After examining its attractors in three environmental contexts for the two-player case, we examine the n-player case.

<u>Two more dynamics</u>. Demonstrating the generality of our approach, we also compute the results for two other types of game agent, those that maximize relative fitness (the relative fitness maximizing agent of evolutionary game theory), and those that maximize group fitness (the relative group fitness maximizing agent of multi-level/group selection theory). We represent the difference by adding social preferences to the self-interested agent, in both pro- and anti-social varieties. Agents driving the self-interested dynamic select games based on whether the equilibrium outcome confers a maximum payoff to them. To define the relative fitness dynamic, we add an antisocial preference, defining the agent to prefer games, all else being equal, whose equilibrium maximizes the difference between their payoffs. In the group selection dynamic, the social preference is prosocial, as agents, all else being equal, select games to maximize the sum of payoffs conferred.

Although these three types of agent—behavioral, evolutionary, and group-selected—are very different from each other, they can be united under a single conceptual umbrella. In environments that have available resources or space, where competition is low, the evolutionary regime that establishes the basis for selective pressure will select for agents who choose games to maximize personal utility (or absolute payoff, as in economic game theory) so that they may expand at the quickest rate. In environments with few free resources and no group structure, agents are instead set up to choose games to maximize relative fitness (or relative payoffs, as in evolutionary game theory). In environments with group structure and few free resources, agents choose games to maximize group fitness (by minimizing relative fitness differences, as permitted by group/multi-level selection and other related theories).

**Measures**

We are overall interested in the attractor games of the various dynamics and how they differ from games in the broader space. Specifically, we are interested in how inequality properties change in attractors, a question that is especially interesting in the self-interested dynamic, which does not prefer either equality or inequality. We offer two measures of equality. One is a space's proportion of "win-win" games, games in which two players share the same top-ranked outcome. Another more sensitive and continuous measure of equality is the GINI coefficient of the payoffs of each game's equilibrium outcome or outcomes.

GINI is a familiar non-parametric equality measure whose values scale between 0 (equal) and 1 (unequal), and that is easily generalizable to discrete payoffs. For an n player game, there are $2^n$ outcomes, each with n payoffs, distributed ranging from 1 to $2^n$ (the number of outcomes to rank). These may be nearly equal to each other or widely varied, a property that GINI can determine. Under this measure, an equilibrium outcome that one player ranks highly, and others rank poorly, will receive a high GINI score close to 1, while an outcome in which all players receive the same payoff (whether all high or all low) will be closer to 0, indicating high equality.

## Results
### Two-player games with absolute fitness maximizing agents

Our motivating questions surround the nature of institutional evolution as driven by self-interested agents. How many attractor institutions are there, how do they differ from the broader space, and, how do the values and features they represent differ from the values of the agents that selected them?

Under the self-interested dynamic, a game is an attractor if it has a unique pure-strategy Nash equilibrium that pays the maximum payoff to the focal player. The attractor games are a subset of the games with exactly one Nash equilibrium. However, not all attractors are win-win and not all win-win games are attractors. For example, the game space includes a representation of the Stag Hunt, which is win-win by our definition but has a second equilibrium that sets it outside of the set of attractors.

Contrary to intuition, "win-win" games are very numerous in the two-player space: one in four games drawn randomly from the two-player ordinal games are win-win (Fig. 3a).

Moving from baseline properties of the space to properties of trajectories over it, we find that 37.5% (54/144) of the two-player games are attractors, and that they form a single contiguous basin of attractors (Fig. 3b). Of games in this basin, 1/2 are win-win, compared to 1/4 of all two-player games. The self-interested dynamic doubles the proportion of win-win games, despite the self-interested agent's absence of social preferences.

### Two-player games with relative- and group- fitness maximizing agents

The proportional doubling of win-win games in attractor institutions holds when agents maximize personal fitness, but this result is sensitive to changes in the type of agent. Given the results of the self-interested dynamic, it is straightforward to compute the attractors of the other two. In the relative group fitness dynamic, the non-win-win games composing half of the attractors become unstable, and the proportion of win-win games becomes 100% (Fig. 3b). In the relative individual fitness dynamic, the opposite happens, leading to a change from 25% win-win games over the whole space, to 0% win-win games in the attractor (Fig. 3b).

### *n*-player games

Generalizing these results to *n* players, we gain further insight into the effects of between-game preferences on institutional evolution.

In general, as game population size *n* increases, the number of two-choice ordinal games quickly becomes astronomical:

$$(2^n!)^n$$

Although the number of 2 player games is less than 1e3, the number of 4 player games is well above order 1e50, and the number of 8 player games is much larger than 1e4000. Fortunately, these vast sums do not seem to undermine the countability of the domain.

### Attractors in the *n*-player games

As the attractors of the evolutionary and group-level dynamics are a subset of those of the self-interested dynamic, we attend first to the properties of the self-interested agent's attractors as *n* increases. We find, numerically, that attractors become a steadily decreasing proportion of games (Fig. 4a).

This might at first glance seem to imply that dynamics become more important in steering evolution toward certain types of attractor games, as the vanishing number of attractors and increasing number of games contribute to an increase in the average number of steps to convergence. However, attendant with these effects of population growth, driven mainly by the increasing dimensionality of constituent games, is an explosion in the number of nearest neighbors per game, which in turn drives an explosion in the number of shortest paths between arbitrary pairs of games, and, ultimately, very unconstrained dynamics.

Given these results for the self-interested agent, the results for the other two dynamics are straightforward and trivial. No new factor resulting from the increase of *n* pulls the relative fitness dynamic from 0%, or the group-level dynamic from 100%. One observation, comparing across the dynamics, is that the self-interested dynamic's crashing equality makes its institutional evolutionary outcomes more difficult to distinguish from the active selection against equality that we observe in the relative fitness dynamic; socially "neutral" behavior is effectively antisocial in large-n games.

**Scaling of inequality in the self-interested dynamic**

Focusing again on the self-interested dynamic, we look more closely at questions of equality. With the explosion in the number of games with increasing *n* comes a crash in the proportion of win-win games (including "win-win-...-win" games, and so on) with size:

$$2^{-n^2}$$

As a fraction of the total number of games, this value decreases super-exponentially from one in four two-player games being win-win, to fewer than one in a billion win-win games at n=6. As a fraction of attractors, which themselves constitute a declining proportion of large-n games, win-win games decline much more quickly, such that the win-win property becomes vanishingly rare even among games that are selecting for them.

However, the appeal of the win-win property as a game-level equality estimator is less about their suitability to the task and more about how easy they are to count. Alternative measures allow more sensitive judgements about the scaling of inequality in attractor institutions as the self-interested dynamic scales to larger populations.

The GINI coefficient is one of these appealing alternatives. We compare the GINI coefficients over the payoffs of Nash outcomes of attractor and non-attractor games (Fig. 4b). The difference between them quickly becomes negligible, consistent with our other findings that, in games with a large number of players, the self-interested dynamics select for games that are only desirable to the single favored agent driving the dynamic.

## Discussion

The interactions that structure our daily lives are not randomly selected from the space of social systems, nor from the small subset of systems, such as the Prisoner's Dilemma and Stag Hunt, whose prominence derives from their ability to illustrate academic points. Institutions and other social structures — languages, rites, and systems of culture — develop through a process that can be conceptualized as a trajectory through institution space. When agents have preferences over games, and the ability to make incremental changes to those games, they can dramatically remake the space of likely institutions. In particular, we find that win-win outcomes change in proportion from 25% to 0%, 50%, or 100%, depending on whether we are considering agents that drift through the games randomly, or agents motivated to maximize relative personal fitness, absolute personal fitness, or relative group fitness, respectively. The 50% case is especially illustrative: self-interested agents converge on a subset of games that are disproportionately fair in the equilibrium outcomes they provide because fairness tends to co-occur with the combination of predictability, stability, and efficiency that they seek out. This particular property seems not to scale with game size: incidental fairness very quickly becomes negligible as populations of self-interested agents grow. From the background of this result, the other two dynamics can be seen as implementing very weak "ceteris paribus" pro- or anti- social preferences.

**An alternative account of norms and conventions**

One contribution of this work is to offer an alternative account of informal institutional constraints like norms, conventions, and other proposed mechanisms for the real-world state of affairs in a society: most of our daily interactions are rote, non-strategic, and even mundane. Existing conceptions of norms, conventions, and other proto-institutional structures understand the emergence of these constructs in terms of regularities in within-game behavior: agents have fixed choices, some combinations of their choices are desirable but unstable, and agents must develop a scheme for making desirable combinations likely despite strategic considerations. Under a strictly within-game framework, the major questions are how the scheme emerges and how it persists. These questions become much more straightforward under a between-game conception of institutional emergence. Our simulations support a picture in which a norm or convention emerges as a result of institutional dynamics that drive agents toward entirely "non-strategic" win-win games (in which players' interests are naturally aligned) or even "non-game" no-conflict games (in which their interests are orthogonal). Given a choice between a game that requires trust in another and one that does not, or one that imposes a conflict of interest and one that does not, it seems natural agents would select the less fraught institution. Within our institutional evolutionary framework, the convergence of a population of agents upon some stable pattern of socially efficient behavior is largely a function of institution-scale processes, rather than strictly behavioral processes. Agents who are subject to a norm or convention have not just converged on one versus another passive pattern of behavior, they actively perceive shifted payoffs, different consequences, and new strategic affordances than those who are not subject, they are *in* a different game and their collective outcome can only be modeled satisfactorily with a between-game formalism.

**The pair as the most common scale of institutional organization**

With general tools, we gain the ability to integrate observations from different disciplines and frameworks under a common umbrella. One classic descriptive finding in anthropology is that persistent mating pairs (i.e. marriages) are an organizing principle common to many human societies (32). Work focused across disciplines and organizational forms has found increasingly large institutions to be increasingly susceptible to elite capture, unfairness and other deviations from win-win ideals (33, 34). While theories of the general phenomenon of capture have focused on active mechanisms, it may be that it falls passively out of how game spaces scale: that in systems with large populations, inequality provides the baseline against which more ideal, unlikely institutions must distinguish themselves.

**A mechanism for major transitions in human evolution**

Comparing outcomes across evolutionary, group-level, and economic agents is especially valuable in the case of species, such as humans. Human evolutionary history has been punctuated by major transitions in which qualitative shifts to new forms of organization have fundamentally altered societies. But explaining these shifts in evolutionary terms is difficult. In one narrative, the innovations of genus *Homo*—tool-assisted hunting, processing of food, and alloparenting —created a new low-competition niche that transitioned human ancestors from population biology's typical evolutionary regime of relative fitness maximization into one of absolute fitness maximization (35). This expansion into open niches has been shown to have unique dynamics, allowing for traits that otherwise would not exist (37). Then as the human niche filled and resource competition increased, humans transitioned into the regime that characterizes human cultural evolution: group-level selection operating on the basis of relative payoff differences.

Allowing this kind of narrative, an evolutionary account of major transitions, fails to alone explain how the subtle differences between, for example, the maximization of relative versus absolute fitness could trigger major organizational changes in a society. However, it is precisely these subtle differences that, in our model of institutional evolution, so dramatically change the makeup of institutional attractors as to alternately magnify, suppress, or guarantee equitable social outcomes. Allowing institutional preferences to drive trajectories through game space provides a mechanism for major transitions, as they amplify differences between otherwise subtle game theoretic distinctions.

**Limitations and future work**

Our results hold for a narrow subset of games, namely those two-choice normal form games with ranked payoffs. However, extensions of this space to more choices or continuous payoffs are not likely to change our findings. Seen as equally-spaced partitions of continuous game spaces, the statistics of the ordinal games should remain proportional to those of the internal regions they define. And increasing the number of choices by two should only make our key findings stronger, particularly the decline of win-win games as population increases. More players means more outcomes, for example four outcomes for two players, eight outcomes for three players, sixteen outcomes for four players, and so forth. Thus, increasing the number of participants and outcomes inherently decreases the proportion of games that benefit all.

The topology also imposes limitations that make it impossible to test other types of preferences that agents are likely to have, such as a preference for extensive over normal form games, repeated versus single-shot relationships, more versus fewer choices, or more versus less information about others' beliefs.

Furthermore, our agents are ultimately artificial, and establishing the generality of our findings will require comparisons to real world institutional preferences. Fortunately, where such artificially selfish agents converge on prosocial outcomes, more realistic agents need only mind pro-social biases to be at least as likely to do the same. Against this promising background, our institutional evolutionary framework makes behavioral studies simple to articulate. In one design we have developed, two participants play a randomly selected game from the topology, and a neighboring game, and are allowed to choose which of the two to play a second time. By repeating this procedure for many game pairs, an investigator can infer the game features that drive participants' preferences between institutions, and directly compute the attractor games that those preferences drive dynamics towards. In fact, we explicitly developed this framework with testability in mind in order to ground the assumptions of this work in behavioral data and test its predictions about the defining features of attractor institutions.

The flexibility of this framework makes it useful for a variety of other important problems. By explicitly modeling game change dynamics, we make it possible to test the effects other important dynamical phenomena on institutional evolution, such as history dependence, emergent diversity, neutral evolution, heterogeneous agents, the interaction of rules with culture, and the coevolution of within-game experiences and between-game preferences. For an example of possible extensions, consider the variety of aggregation rules. In the dynamics we consider here, the "winner" of a specific game outcome gains unilateral control over the next choice of game, a choice that may enable them to ensure that they continue to win. But in simple variations, the choice of game could be driven by the choice of a randomly selected agent, or, in a model of complex collective action, the preferences of a majority or plurality of players.

## Conclusion

Agents in a mutable social environment can change their incentives as those incentives change them. These institutional change processes are of fundamental interest to both evolutionary and behavioral game theory in general, and institutional analysis in particular. Still, we have been lacking tractable frameworks for representing the richness of institutional evolution. Dynamics over the topology of games offer a parsimonious representation of institutional evolutionary processes. Within this framework, institutional change is a trajectory over neighboring games in which players evolve the games they play in by incrementally making their payoff structures more favorable. Preferences can take into account many qualities, but we focus on three simple, selfish qualities: predictability of outcomes, existence and uniqueness of Nash stable outcomes, and efficiency of outcomes. We find that the nature of "institution space" can impose constraints that encourage socially beneficial outcomes even among agents with no interest in those outcomes, but only in small social systems.

We advance a view that humans and other animals are not caged subjects of immutable institutions. Institutions evolve, often due to pressures exerted by their participants. The games we encounter are themselves the endpoints of dynamics that select and replicate certain structures over others. Elucidating the properties of "attractor" institutions sheds light on the emergence of organized human groups

## Acknowledgements

The authors wish to thank Bryan Bruns, Austin Shapiro, Pete Richerson, Monique Borgerhoff Mulder, Cristina Moya, and the EEHBC group at University of California, Davis. Author SF conceived of the research, computed the numerics, and wrote the manuscript. Author CA conceived of the research and contributed to the manuscript. This work was supported in part by the Neukom Institute for Computational Research.

**FIGURES**

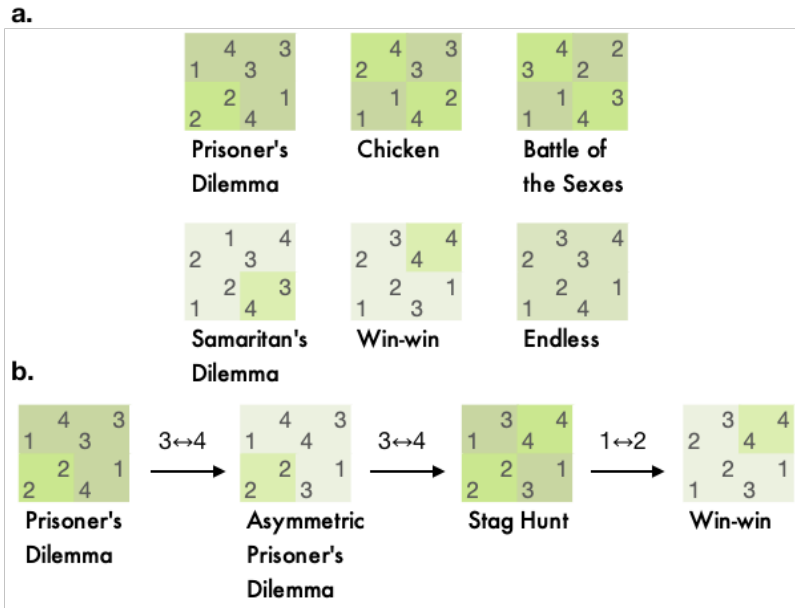

**Fig 1. A sampling of ordinal games, with a trajectory through them.**
 **a.** The space of 2-player, 2-choice ordinal games includes a variety of interesting and relevant games, including well-studied games such as the Prisoner's Dilemma, Chicken, and Battle of the Sexes, and less remarked upon games such as win-win games that don't require strategy, cyclic games without pure-strategy equilibria, and asymmetric games. Outcomes that are Nash equilibria are slightly brighter. By ordinal, we mean games with consecutive integer payoffs up to the number of outcomes. In these illustrations, 4 is high and 1 is low.  Ordinal games have the advantage of being amenable to counting.
 **b.** Two games in this space are neighbors if they differ by a swap of similar payoffs. Assuming that most institutional change is incremental, institutional evolution can be modeled as a trajectory through the neighboring games. Here we illustrate how an agent might incrementally evolve a prisoner's dilemma into a win-win game. This trajectory terminates on the game "Win-win", which is an attractor for self-interested agents who prefer stable, predictable, and efficient games (defined herein as offering a unique pure-strategy Nash equilibrium that confers a maximum payoff).  Example adapted from (25).

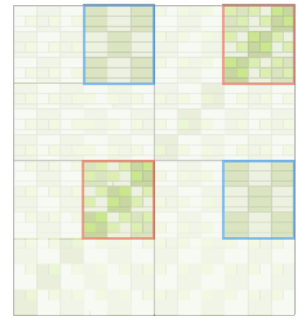
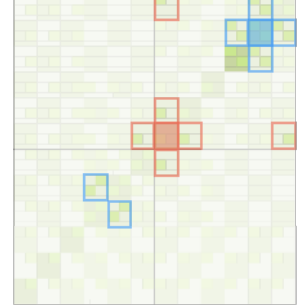
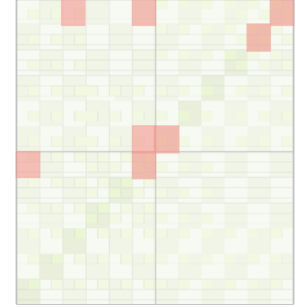

**Fig 2 The space of two-player games, with masks illustrating some of its properties.**

  **a.** A simple representation of the space of 144 two-player, two-choice games with ordinally ranked payoffs. Observe that symmetric games, occupying the increasing diagonal, are a minority, and that the lower-left quarter of games are win-win, in the sense of having one outcome that confers the maximum payoff of 4 to both players (also see Fig 3a). Spaces with more than two players are much larger and more difficult to diagram than this.

  **b.** This mask of panel **a.** illustrates the Nash properties of games in this space. The games in the blue outlines have no pure strategy Nash equilibria. The games in red outlines have two. The remaining games, 75%, have exactly one pure strategy Nash equilibrium. We discuss how this distribution changes as the number of players increases.

  **c.** This mask of **a.** illustrates the complex nature of neighbor relations in the two-player space. The red outlines show the six "neighbors" of the prisoner's dilemma. The blue outlines show the neighbors of the Battle of the Sexes. Note that some adjacent games are not visually adjacent, a shortcoming of the grid representation of what is in truth a much more complex topology.

  **d.** This mask of **a.** shows the locations of the games in Figs. 1a and 1b.

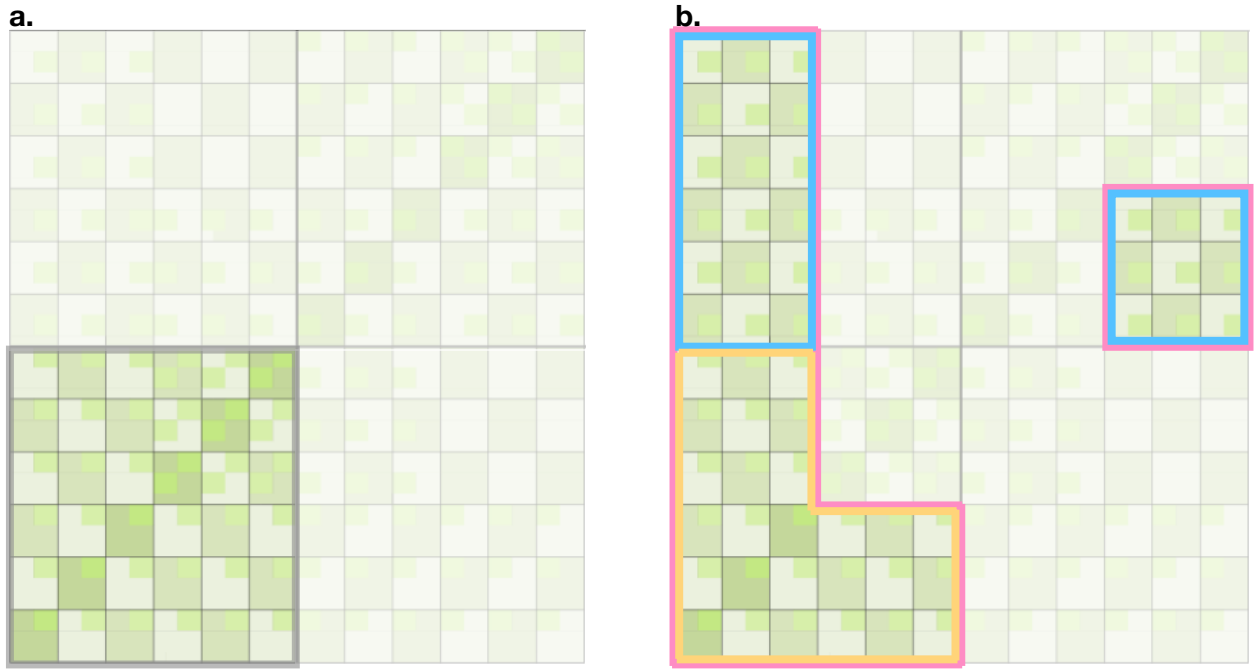

**Fig 3. The win-win games and the attractor games of three institutional dynamics.**

   **a.** Win-win games account for 1/4 of the games in the two-player space, shown here in the lower-left quadrant. See Fig 2a for details of each, namely that each has an outcome conferring the maximum payoff of 4 to both players.

   **b.** The encompassing pink outlines shows the basin of attractors that results from self-interested agents' evolutionary trajectories. Note that these attractor games form a contiguous block: as the 9 games on the right have several neighbors among the games in the block on the left, via swaps that are not apparent from this grid representation. Note also that half of these attractor games are in the win-win quadrant. Compared to panel a, the institutional evolutionary process double the chances of converging upon a win-win game, even though the selfish agents driving it have no explicit preferences for mutually beneficial games. With the attractors of the multilevel selection dynamic (orange outline), that probability increases to 100%, as all of its attractor games are win-win. Conversely, the fitness maximizing evolutionary agent will converge on the more unfair games in the upper half (green outline), all of which have an asymmetry between the focal player's earnings and those of the other player.

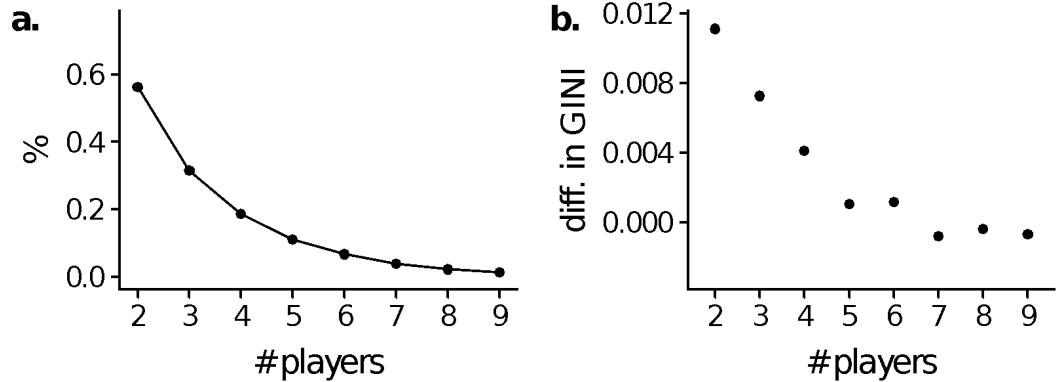

**Fig 4. Emergent fairness driven by the self-interested dynamic disappears as player populations grow.**

**a.** Attractors become a smaller fraction of games with increasing population. The black line, derived from simulation, gives the computed proportion of games up to 9 players that are attractors of the self-interested dynamic. **b.** Each game outcome contains payoffs for each player, and those payoffs can differ widely from each other. We compute the GINI coefficient of the payoffs in Nash outcomes, and compare them within the attractors and in the full space. We find that the difference quickly becomes negligible, telling a complementary story to that of Fig. 3b, that the self-interested dynamic biases institutional evolutionary processes to emergently select fair games when populations are small, an effect that disappears for larger populations.

# SUPPORTING FIGURES

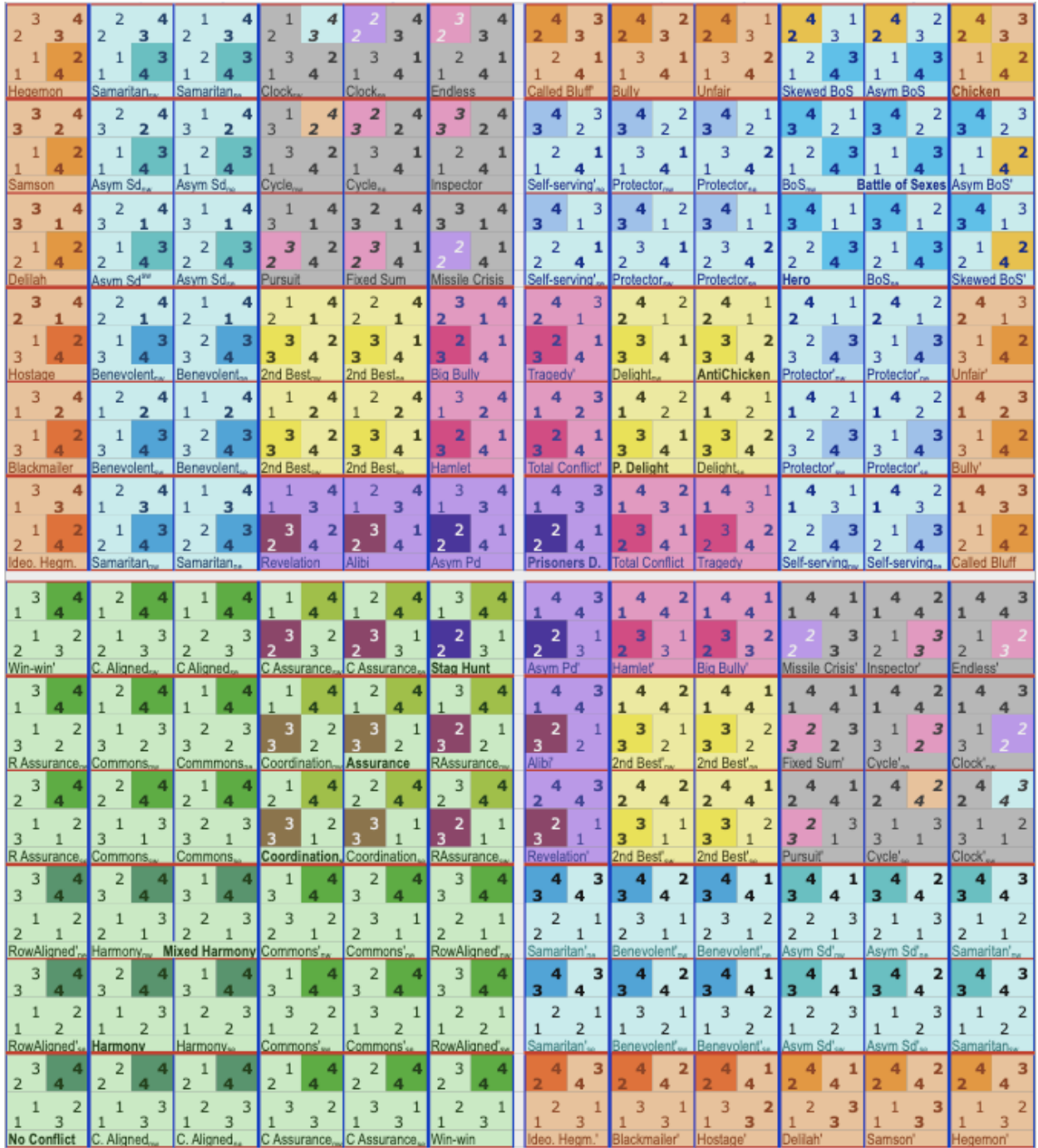

**Fig S1. Richer visual representation of the topology of 2-player, 2-choice ordinal games.** This figure, an elaboration of Fig. 2a, illustrates neighboring relations in the blue, red, and quadrant boundaries, and game classes by background color. Copied with permission from (23).